\documentclass[aps,prd,showpacs,preprintnumbers,nofootinbib]{revtex4}
\usepackage[dvips]{graphicx}
\usepackage{bm,latexsym,amsmath,amssymb,amsfonts}

\begin{document}

\title{Multipole moments for black objects in five dimensions}

\author{Kentaro Tanabe}
\affiliation{Yukawa Institute for Theoretical Physics, Kyoto University, Kyoto 606-8502, Japan}
\author{Seiju Ohashi and Tetsuya Shiromizu}
\affiliation{Department of Physics, Kyoto University, Kyoto 606-8502, Japan}
\begin{abstract}
In higher dimensions than four, conventional uniqueness theorem in asymptotically flat 
space-times does not hold, i.e., black objects can not be classified only 
by the mass, angular momentum and charge. In this paper, we define multipole 
moments for black objects and show that Myers-Perry 
black hole and black ring can be distinguished by quadrupole moments. This 
consideration gives us a new insight for the uniqueness theorem for black objects 
in higher dimensions.
\end{abstract}


\maketitle

\section{Introduction}

In four dimensions, stationary and asymptotically flat black hole solutions 
can be classified by their mass, angular momentum and charge completely. 
This is the famous uniqueness theorem \cite{Israel}. 
On the other hand, this uniqueness property of black objects does not hold in higher 
dimensions. As presented by Emparan \& Reall \cite{Emparan:2001wk,Emparan:2001wn,Emparan:2008eg}
(See also Ref. \cite{Pomeransky:2006bd}), in five dimensions, there is the black ring solution, which can 
have same mass and angular momentum as the Myers-Perry black hole \cite{Myers:1986un}. 
If we do not restrict our consideration to cases with single horizon, 
there are many, probably infinite, regular solutions 
with same mass and angular momentum \cite{Elvang:2007rd,Evslin:2007fv,Iguchi:2007is,
Izumi:2007qx,Elvang:2007hs}. This shows that there are much richer properties of black object 
solutions in higher dimensions compared to four dimensions. At the same time, however, 
it is unlikely that the complete classifications of these black 
objects are possible. Now, note that there are a sort of uniqueness theorems in some 
restricted cases. For example, in static 
and vacuum space-times, the Schwarzschild-Tangherlini solution \cite{Tangherlini:1963bw} is only 
regular black hole solution \cite{GIS}. For stationary solutions which have single horizon and 
two axial commuting Killing vectors, if one specifies the topology of the horizon 
($S^3$ or $S^{1}\times S^2$), the solution can be uniquely determined (the Myers-Perry solutions or 
black ring solutions, respectively) \cite{Morisawa:2004tc,Morisawa:2007di,Rogatko:2008yd}. 
Furthermore, in more general situations of five dimensional, stationary 
and two rotational symmetric asymptotically flat space-times, if one specifies its mass, 
angular momentum and so called rod structure \cite{Emparan:2001wk,Harmark:2004rm}, 
which represents the positions of event horizons and rotational axis, the regular solution 
is determined uniquely \cite{Hollands:2007aj}. Theorems like these can be extended to 
non-vacuum cases and so on \cite{Hollands:2007qf,Hollands:2008fm,Tomizawa:2009ua,Tomizawa:2009tb,
Armas:2009dd,Tomizawa:2010xj,Emparan:2010ni}. 
However, these uniqueness theorem for five dimensional stationary 
black objects are not satisfactory in physical point of view. This is because 
the relation between the rod structure and global charge is unclear and we want 
to classify black object space-times in terms of global charges or quantities observed at 
infinity. Unfortunately, the rod structure is quasi-local concept. 
The purpose of this paper is the classifications of 
stationary black objects by multipole moments, which are defined at spatial infinity. 
The asymptotic quantities like multipole moments might be useful to 
study the properties of solutions(black ring solutions 
in $d>5$ dimensions \cite{Emparan:2007wm} or non-Myers-Perry black hole with spherical 
topology of event horizon \cite{Dias:2009iu,Dias:2010eu}) which are conjectured to be exist.   

Geroch \cite{Geroch:1970cc,Geroch:1970cd} and Hansen \cite{Hansen} 
defined the multipole moments by using the conformal completion to obtain the property 
of space-times at spatial infinity $\Lambda$ in four dimensions. In Ref. \cite{Geroch:1970cd}, Geroch 
conjectured that (A) two solutions of the four dimensional 
Einstein equations having the same multipole 
moments coincide each others at least in a neighborhood of $\Lambda$, and (B) given any sets 
of multipole moments, subject to the appropriate convergence condition, there exist 
a solution of Einstein equations having precisely those moments. About the conjecture 
(A), Beig \& Simon and Kundu showed the validity for static \cite{Beig:1972} and 
stationary space-times \cite{Beig:1981,Kundu:1980rn}. For the conjecture (B), there 
are not rigorous proof and it is open issue even in four dimensions still now. There is also 
the coordinate based definition of multipole moments by Thorne \cite{Thorne:1980ru}. 
It was shown that Thorne's multipole moments are same as Geroch and Hansen's multipole 
moments under certain conditions \cite{Gursel}. Following Geroch's idea on four dimensional 
static cases, Tomizawa and one of the present authors proposed the definition of 
multipole moments in higher dimensional {\it static} space-times \cite{Shiromizu:2004jt}. 
In this paper, we discuss the definition of multipole moments 
in five dimensional {\it stationary} space-times and then show that asymptotically flat, 
stationary and two rotational symmetric solutions with {\it single} horizon are completely classified 
by the mass monopole, quadrupole moments and angular dipole moments. This successful result 
will encourage us to study the classification for general cases including multiple horizon cases.  

The rest of this paper is organized as follows. In the next section, we define mass multipole 
moments and angular multipole moments in five dimensional stationary space-times. 
We will emphasize that the definition of the angular multipole 
moments is rather non-trivial task. Some details related to the definitions is shown in appendix A. 
In Sec. III, as an exercise, we shall consider the multipole moments of the static black 
objects and discuss the classification of them. In Sec. IV, we will compute 
the multipole moments for stationary black objects. Then we show that black ring can be distinguished 
from the Myers-Perry solutions by the ``reduced" quadrupole moments which are well defined in 
center-of-mass gauge. In Sec. V, we summarize our result and discuss the possibility of uniqueness 
theorem using the multipole moments. In appendix A, we write down the field equations and see 
why our definition of multipole moments is appropriate. In appendix B, we compute the multipole moments 
for the black ring solutions with two angular momenta. In appendix C, for comparison, we compute 
the multipole moments for black objects with multiple horizons solutions like black Saturn 
\cite{Elvang:2007rd} and orthogonal black di-ring solutions \cite{Izumi:2007qx}.

\section{Definition of multipole moments}

In this section, following Refs. \cite{Geroch:1970cd,Shiromizu:2004jt}, we shall define 
the multipole moments in five dimensional stationary space-times. At first, we describe the 
definition of the asymptotically flatness based on the conformal completion method briefly. 
Then we will give a definition of the multipole moments. The definition of the 
multipole moments associated with angular momentum is not given by a simple extension 
from four to five dimensions. We also address the gauge dependence which comes from 
the gauge freedom of the conformal transformation.

\subsection{Asymptotic flatness}

For stationary space-times, we introduce the notion of asymptotic flatness at spatial infinity
based on conformal completion method \cite{Geroch:1970,Geroch:1977jn}. The metric of 
stationary space-times can be written as
%
\begin{equation}
\hat{g}_{ab}\,=\frac{1}{\lambda}\,\xi_{a}\xi_{b}+\hat{h}_{ab},   \label{metric}
\end{equation} 
%
where $\xi=\partial/\partial t $ is the timelike Killing vector, $\lambda=\hat{g}_{ab}
\xi^{a}\xi^{b}$ and $\hat{h}_{ab}$ is the metric on $t=\text{const}.$ hypersurfaces. 
Since the multipole moments will be defined on $t=\text{const}.$ hypersurfaces, we 
can focus on only the metric on $t=\text{const}.$ hypersurfaces. 

Let us consider the conformal transformation as 
%
\begin{equation}
h_{ab}\,=\,\Omega^{2}\hat{h}_{ab} \label{conf-trans} .
\end{equation}
%
If there is a function $\Omega$ which satisfies the conditions
%
\begin{equation}
\Omega\hat{=}0\, , \, D_{a}\Omega\hat{=}0\, , \,D_{a}D_{b}\Omega\,\hat{=}\,2h_{ab},
\end{equation}
%
$t=\text{const}.$ hypersurface is called asymptotically flat space 
and the point $\Omega=0$ is identified as the spatial infinity $\Lambda$. Here, $D_{a}$ is 
the covariant derivative with respect to the metric $h_{ab}$ and $\hat{=}$ stands for the 
evaluation on $\Lambda$. At the spatial infinity,  $h_{ab}$ becomes the flat metric
and we use the coordinate as follows 
%
\begin{equation}
ds^{2}\,=\,h_{ab}dx^{a}dx^{b}
\,\hat{=}\,d\rho^{2}+\rho^{2}(d\theta^{2}+\sin^2\theta d\phi^2+\cos^{2}\theta d\psi^{2}) \label{bgmetric}.
\end{equation}
%
The above conditions on $\Omega$ imply the asymptotic behavior of $\Omega$ as 
$\Omega\sim1/r^2\sim \rho^2$ near $\Lambda$.
In this paper, we consider only the solutions of the vacuum Einstein equations 
$\hat{R}_{ab}^{(5)}\,=\,0$ for simplicity. 

\subsection{Multipole moments}

At first, we define mass $2^{s}$-pole moments $P_{a_{1}a_{2}\cdots a_{s}}$ as 
%
\begin{gather}
P\,\equiv \,\Omega^{-1}(1-\sqrt{\lambda}) \\
P_{a}\,=\,D_{a}P \\
P_{a_{1}a_{2}\dots a_{s}}\,=\,\mathcal{O}\left[D_{a_{1}}P_{a_{2}\dots a_{s}} -\frac{(s-1)^2}{2}
R_{a_{1}a_{2}}P_{a_{3}\dots a_{s}} \right], 
\end{gather} 
%
where $\mathcal{O}[T_{ab\cdots}]$ denotes the totally symmetric trace free part 
of the tensor $T_{ab\cdots}$. In four dimensions, the mass multipole 
moments in stationary space-times are defined from the scalar potential $P$ which consists 
of the lapse function $\lambda$ and the twist potential $\sigma$ satisfying 
a conformal invariant equation \cite{Hansen}. As seen soon, the corresponding twist potential 
is represented by a 
vector $\sigma_{a}$, not a scalar, in five dimensions. 
Hence, we 
use only the lapse functions $\lambda$ for the definition in five dimensions, 
and this does not matter. 
In fact, we can check that the twist potential 
does not contribute to the multipole moments in four 
and five dimensions. The role of $\sigma$ will be important only in the proof of the smoothness 
of the multipole 
moments in four dimensions \cite{Hansen,Beig:1981,Kundu:1980rn}. Note that 
the relation between the Arnowitt-Deser-Misner (ADM) mass and the mass 
monopole is shown in Ref. \cite{Harmark:2004rm}
%
\begin{equation}
M_{\rm ADM}\,=\,\frac{3\pi}{4}P. 
\end{equation}
%
 
For the definition of angular multipole moments, we assume the presence of the two commuting 
axisymmetric Killing vectors $m=\partial/\partial\phi$ and $l=\partial/\partial\psi$. 
In this case, there are many known exact solutions. However, 
note that this assumption might be rather strong in some senses. 
This is because the existence of only single axisymmetric Killing vector is guaranteed from the stationarity
\cite{Hollands:2006rj,Hollands:2008wn}. Although it is interesting to consider the multipole 
moments of 
single rotational symmetric cases too \cite{Reall:2002bh}, this is beyond the scope of our current paper. 

Now, we introduce the tensor $\hat{\sigma}_{ab}$ as
%
\begin{equation}
\hat{\sigma}_{ab}\,=\,\hat{\varepsilon}_{abcde}\xi^{c}\hat{\nabla}^{d}\xi^{e}.
\end{equation} 
%
Since $\hat{\sigma}_{ab}$ satisfies $\hat{D}_{[a}\hat{\sigma}_{bc]}=0$ by the vacuum Einstein 
equation ${}^{(5)}\hat{R}_{ab}=0$, $\hat{\sigma}_{ab}$ can be written by the twist potential 
$\hat{\sigma}_{a}$ as 
\begin{equation}
\hat{\sigma}_{ab}=\hat{D}_{[a}\hat{\sigma}_{b]}.
\end{equation}
$\hat{\sigma}_{a}$ satisfies the four dimensional Maxwell-type equation (see appendix A). 
Therefore, under the conformal 
transformation of Eq. (\ref{conf-trans}), we can take $\hat{\sigma}_{ab}$ as conformal 
invariant quantity, i.e., $\hat{\sigma}_{ab}\,=\,\sigma_{ab}$ and 
$\hat{\sigma}_{a}=\sigma_{a}$. 
From this twist potential $\sigma_{a}$, we can construct two scalar potentials as    
%
\begin{equation}
J^{\phi}=\frac{\sigma _{a}l^{a}}{(l_{a}l^{a})^{1/2}}~~{\rm and}~~
J^{\psi}=\frac{\sigma _{a}m^{a}}{(m_{a}m^{a})^{1/2}}.
\end{equation} 
%
Then we can define the angular multipole moments in the same way as the mass multipole 
moments discussed above. 
That is, the angular $2^{s}$-pole moments $J^{\phi}_{a_{1}a_{2}\cdots a_{s}}$ 
and $J^{\psi}_{a_{1}a_{2}\cdots a_{s}}$ are defined recursively as
%
\begin{equation}
J^{\phi}_{a_{1}a_{2}\cdots a_{s}}=\mathcal{O}\left[D_{a_{1}}J^{\phi}_{a_{2}\dots a_{s}} 
-\frac{(s-1)^2}{2}R_{a_{1}a_{2}}J^{\phi}_{a_{3}\cdots a_{s}} \right] 
\end{equation} 
%
and
%
\begin{equation}
J^{\psi}_{a_{1}a_{2}\cdots a_{s}}=\mathcal{O}\left[D_{a_{1}}J^{\psi}_{a_{2}\dots a_{s}} 
-\frac{(s-1)^2}{2}R_{a_{1}a_{2}}J^{\psi}_{a_{3}\cdots a_{s}} \right] .
\end{equation} 
%
Here we have a remark: if one considers the cases having single rotational symmetry, 
$\sigma_{a}$ cannot be written only by scalar potential in a natural way. 
This means that it is not 
easy to construct the angular multipole moments in single rotational symmetric cases. 
The resolution to this difficulty is left for future study. 

\subsection{Unphysical gauge dependence}

Before computing the multipole moments for known black objects, we should comment on the 
gauge dependence of the multipole moments defined above. There are gauge freedoms 
in the conformal completion of Eq. (\ref{conf-trans}) as
%
\begin{equation}
\Omega \rightarrow \omega\Omega, \label{gauge-trans}
\end{equation} 
%
where 
%
\begin{equation}
\omega \simeq 1 +\frac{f(\theta,\phi,\dots)}{r}+O(1/r^2)\label{omega}.
\end{equation} 
%
Under this gauge transformations, multipole moments are transformed as
%
\begin{equation}
P_{a_{1}\dots a_{s}} \rightarrow P_{a_{1}\dots a_{s}} - s^{2} \mathcal{O}
\left[ P_{a_{1}\dots a_{s-1}}D_{a_{s}}\omega\right] \label{multi-trans}
\end{equation} 
%
in the linear order of $\omega$
\footnote{Transformations of $J^{\phi}_{a_{1}\dots a_{s}}$ and $J^{\psi}_{a_{1}\dots a_{s}}$ 
are same as Eq. (\ref{multi-trans})}.
$D_{a}\omega$ represents the $1/r$-order part 
in $\omega$ (See Eq. (\ref{omega})) and corresponds to the choice of the origin of the 
coordinate in 
the physical coordinate $\hat{x}^{a}$ \cite{Geroch:1970}. In the definition of 
Thorne's multipole moments \cite{Thorne:1980ru}, there is such gauge freedom, 
which is just a translation. Thus, the freedom of the order of $D_{a}\omega$ can be 
fixed by gauge conditions just like a center-of-mass gauge. 
Higher order parts $O(1/r^2)$ in Eq. (\ref{omega}), which can be 
written as $D_{a}D_{b}\omega, D_{a}D_{b}D_{c}\omega,\cdots$, do not contribute 
to the transformation of multipole moments in the linear order. 
On the other hand, in non-linear order of $\omega$, 
the changes of multipole moments under transformations of Eq. (\ref{gauge-trans}) depend
on not only $D_{a}\omega$, but also higher order terms. For example, the octupole moments
are transformed as
%
\begin{equation}
P_{abc} \rightarrow P_{abc} - 9 \mathcal{O}\left[P_{ab}D_{c}\omega \right]
+g\,\mathcal{O}\left[D_{a}\omega D_{b}D_{c}\omega \right] P+\cdots ,
\end{equation} 
%
where $g$ is a numerical constant. Hence the value of the octupole or higher-pole moments 
depend on the choice of $O(1/r^2)$ parts in $\omega$, while monopole, dipole and 
quadrupole moments depend only on $D_{a}\omega$. 
Higher multipole moments than quadrupole, that is, octupole and $2^{4}$-pole moments, have
gauge ambiguities from the conformal transformation even in center-of-mass gauge. 
Since there is no tractable way to fix the gauge freedom of 
$O(1/r^2)$ parts in $\omega$, we will focus on the computation of only monopole, dipole 
and quadrupole moments for known black objects solutions in the center-of-mass gauge.

\subsection{Modes}

In the practical calculations of the multipole moments, we must specify the concrete coordinate 
and modes. In this paper, we use the the coordinate in Eq. (\ref{bgmetric}). 
Then it is easy to see that the quadrupole moments have nine modes as 
%
\begin{gather}
\cos 2\theta, \notag\\
\sin^{2}\theta\sin 2\phi,\sin^{2}\theta\cos 2\phi,\cos^{2}\theta\sin 2\psi,\cos^{2}\theta\cos 2\psi \\
\cos\theta\sin\theta\sin\phi\sin\psi,\cos\theta\sin\theta\sin\phi\cos\psi,
\cos\theta\sin\theta\cos\phi\sin\psi,\cos\theta\sin\theta\cos\phi\cos\psi .\notag
\end{gather} 
%
In two rotational symmetric cases, non-vanishing quadrupole moment mode is only one mode 
of $\cos 2\theta$ in center-of-mass gauge. 
In this paper, then, we define the coefficients of this mode in $P_{\rho\rho}$ as mass quadrupole 
moments $Q$
\footnote{If we define the quadrupole moment as the coefficient in other components of $P_{ab}$, 
the difference will be only sign.}.

\section{Static cases}

In this section, as a first step, we compute the multipole moments for known static solutions. 
Then we will discuss the classification 
of space-times with single horizon. Following the uniqueness theorem \cite{GIS}, 
regular static black object solutions of the vacuum Einstein equation are completely 
classified by its mass, that is, the mass monopole. Other solutions like static black ring have conical singularities and 
they are not regular solutions. However, by computing the multipole moments of these non-regular 
solutions, we can study the dependence of the multipole moments on the topology of horizon. 
Since all angular multipole moments vanish in static cases, 
we will consider only mass multipole moments. This section will be helpful to study the multipole 
moments for stationary 
cases in which we are interested more.

The metric of static and two rotational symmetric space-times in five dimensions can be written in the Weyl 
coordinate \cite{Emparan:2001wk}
%
\begin{equation}
ds^{2}\,=\,-e^{2U_{t}}dt^{2}+e^{2U_{\phi}}d\phi^{2}+e^{2U_{\psi}}d\psi^{2} +e^{2\nu}(dR^{2}+dz^{2}),
\end{equation} 
%
where $U_{t}+U_{\phi}+U_{\psi}=\log R$. 
The solutions are represented by the rod structure which is composed of the zero points of $g_{tt}$, 
$g_{\phi\phi}$ and $g_{\psi\psi}$ and stand for the positions of event horizons and rotational axis.    
For computation of the multipole moments, it is better to use new coordinate given by 
%
\begin{equation}
R=\frac{1}{2}r^{2}\sin 2\theta 
\end{equation} 
%
and
%
\begin{equation}
z=\frac{1}{2}r^{2}\cos 2\theta,
\end{equation} 
%
and take the conformal factor as $\Omega=1/r^2=\rho^2$. 
The coordinate $t, r, \theta, \phi$ and $\psi$ here are same as those in Eq. (\ref{bgmetric}). 
The Weyl coordinate has the gauge freedom
$z\rightarrow z+\text{constant}$ and this gauge freedom 
corresponds to $r\rightarrow r(1+O(1/r^2))$ in the coordinate of Eq. (\ref{bgmetric}). 
As mentioned in the previous section, the monopole, dipole 
and quadrupole moments we compute in the following are independent on this $O(1/r^2)$ order 
gauge transformations. This implies that those moments should be written by the difference 
$a_{i}-a_{j}$ as seen soon later. 
    
\subsection{Schwarzschild black holes}

%
\begin{figure}[tbp]
 \begin{tabular}{cc}
 \begin{minipage}[t]{0.45\hsize}
  \begin{center}
   \includegraphics[width=\hsize,clip]{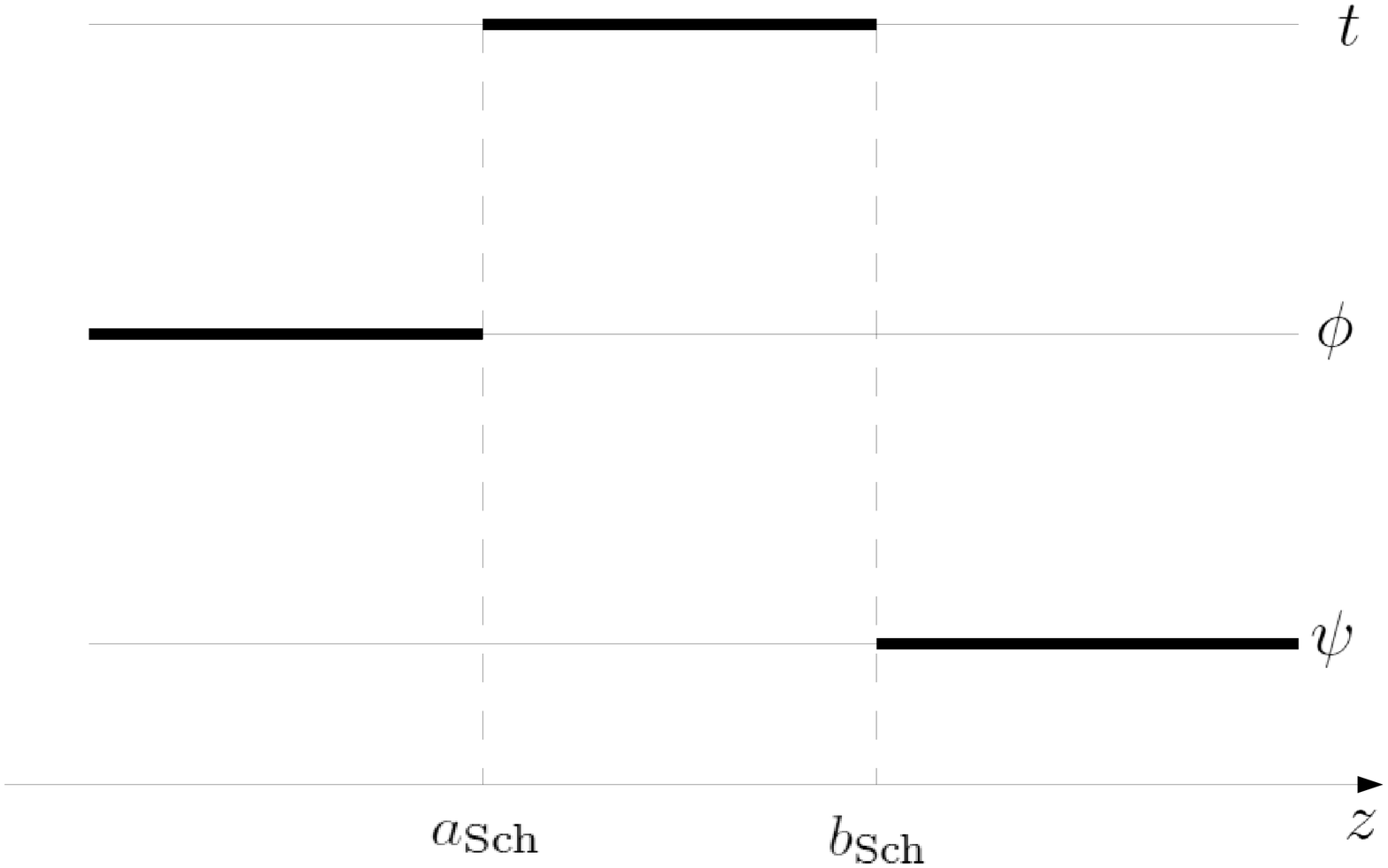}
   \caption{Rod structure of Schwarzschild black hole} 
   \label{fig:one}
  \end{center}
  \end{minipage}
 \begin{minipage}[t]{0.45\hsize}
  \begin{center}
   \includegraphics[width=\hsize,clip]{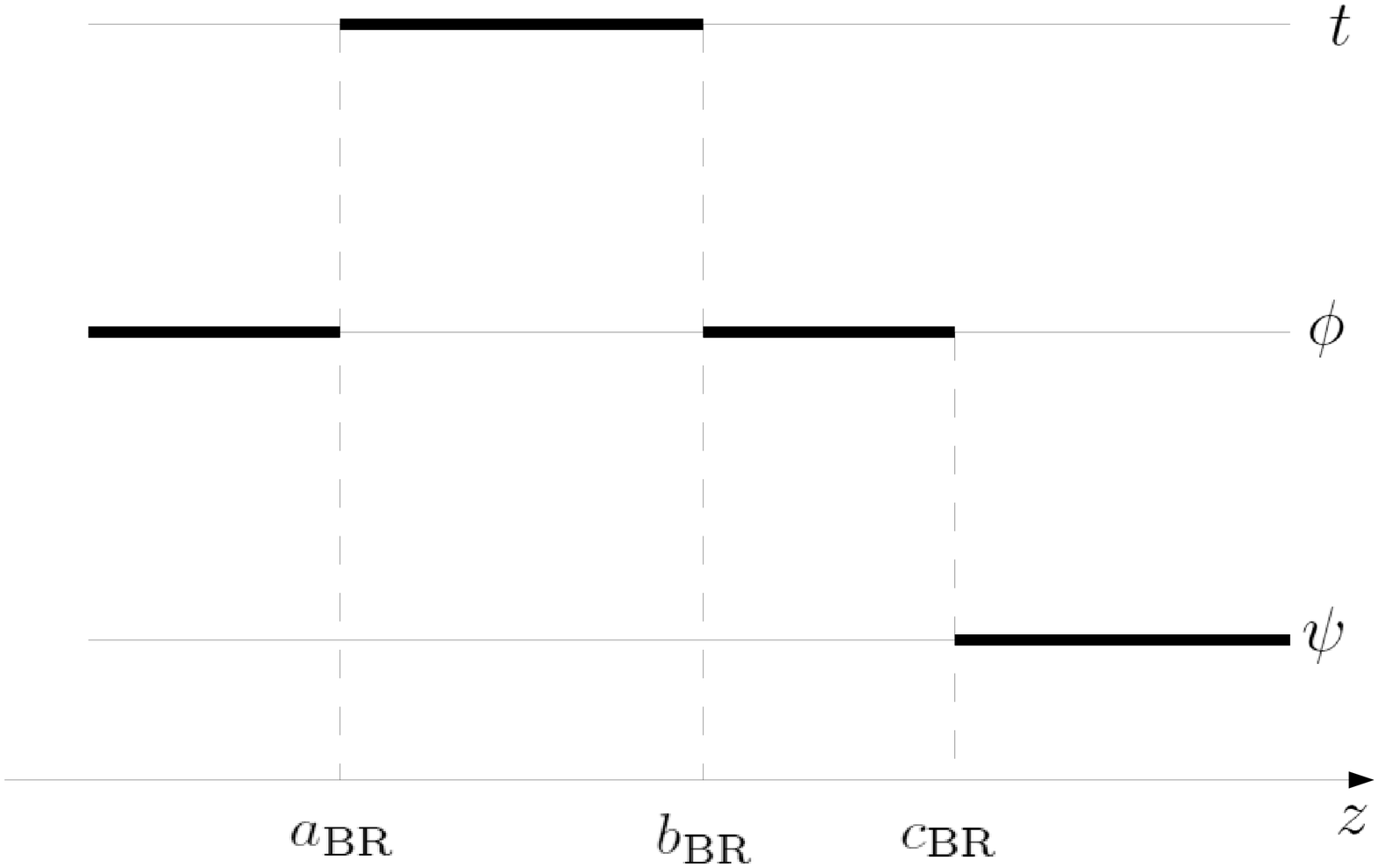}
   \caption{Rod structure of static black ring}
   \label{fig:two}
  \end{center}
  \end{minipage}
 \end{tabular}
\end{figure}
%

At first, we will compute the multipole moments of the Schwarzschild black hole as a trivial example. 
The rod structure of the Schwarzschild black holes are shown in Fig. \ref{fig:one}. The 
Schwarzschild black hole is described 
by two parameters $a_{\text{Sch}}$ and $b_{\text{Sch}}$ in the Weyl coordinate. As there is a gauge freedom 
$z\rightarrow z+\text{constant}$, however, the independent parameter is only 
$b_{\text{Sch}}-a_{\text{Sch}}$. 
After all, the multipole moments of the Schwarzschild black hole are computed as  
%
\begin{gather}
P\,=\,b_{\text{\tiny Sch}}-a_{\text{\tiny Sch}}\,,\,P_{a}\,=\,0 \\
Q\,=\,0. 
\end{gather} 
%
Hence, the Schwarzschild black hole has only monopole as non-trivial multipole moments, 
which is proportional to the ADM mass. Here note that the $2^{4}$-pole 
moment $L$ is given by $L\propto (a_{\text{Sch}}^{3}-b_{\text{Sch}}^3{})$. 
As we stressed before, however, 
it has the unphysical gauge dependence and then the physical meaning of this $L$ is unclear. 

\subsection{Static black ring}

Next, we compute the multipole moments of static black ring solution (the rod structure is 
shown in Fig. \ref{fig:two}). Static black ring solutions have two independent parameters 
$b_{\text{\tiny BR}}-a_{\text{\tiny BR}}$ and 
$c_{\text{\tiny BR}}-b_{\text{\tiny BR}}$. Then, after short calculation, we can see that the 
multipole moments are 
%
\begin{gather}
P\,=\,b_{\text{\tiny BR}}-a_{\text{\tiny BR}}\,,\,P_{a}\,=\,0 \\
Q\,=\,-4(b_{\text{\tiny BR}}-a_{\text{\tiny BR}})
(c_{\text{\tiny BR}}-b_{\text{\tiny BR}}). 
\end{gather} 
%
In our definition of quadrupole moments, $Q$ is always non-positive. In addition, only if 
we take the Schwarzschild limit of $b_{\text{\tiny BR}}=c_{\text{\tiny BR}}$ or flat limit of 
$a_{\text{\tiny BR}}=b_{\text{\tiny BR}}$, the quadrupole moment vanishes. 

\subsection{Classification issue}

We can distinguish the Schwarzschild black hole from black ring solutions by quadrupole moments. Then,  
if the single horizon is assumed, black objects are completely classified by the mass monopole and 
quadrupole moments, which are well defined in center-of-mass gauge. 

Although the cases with multiple horizons is beyond of our current consideration, we have some comments 
on that. In appendix C, we computed the multipole moments for black Saturn and orthogonal
black di-ring cases as examples. 
Static black Saturn and orthogonal di-ring solutions have three and four independent parameters 
respectively. On the other hand, the mass monopole and quadrupole moments can determine only two independent 
parameters. Then the multipole moments up to the quadrupole 
moments are not enough 
parameters to specify the space-times uniquely. That is, higher multipole moments are needed for 
the classification of these solutions. 
As mentioned in previous section, however, the higher multipole 
moments have the unphysical gauge ambiguities of $\omega$. Thus, we should fix the gauge about 
the conformal factor $\Omega$ completely or improve the definition of the higher multipole moments.  

\section{Stationary cases}

In static cases, we have shown that black objects with single horizon can be classified 
by mass monopole and quadrupole moments completely. In this section, we consider the 
black objects with angular momentum and single horizon. As 
we will see later soon, in stationary cases, 
rotating black objects can be classified by mass monopole, quadrupole and angular dipole 
moments. 

Using the fact of $\sigma_{\phi}\sim -\tan^2\theta g_
{t\psi}$ and $\sigma_{\psi}\sim \cot^2\theta g_{t\phi}$ near the spatial infinity, we define
the coefficient of 
$\cos\theta$ in $J^{\phi}_{\rho}$ and $\sin\theta$ in $J^{\psi}_{\rho}$ as angular 
dipole moments $J_{\phi}$ and $J_{\psi}$. Here note that $\cos\theta$ is $l=1$ mode of scalar 
harmonics in the $\phi$-rotational plane with the metric $d\theta^{2}+\sin^{2}\theta d\phi^{2}$ and 
$\sin\theta$ is one in the $\psi$-rotational plane with the metric $d\theta^{2}+\cos^{2}\theta d\psi^{2}$. 
Note that the relations between the angular dipole moments and the 
ADM angular momentum are given by \cite{Harmark:2004rm}   
%
\begin{equation}
J^{\phi}_{ADM}\,=\,\frac{\pi}{4}J_{\phi}\,,\,J^{\psi}_{ADM}\,=\,-\frac{\pi}{4}J_{\psi}.
\end{equation} 
%

\subsection{Myers-Perry black holes}

Let us examine the Myers-Perry solutions. The metric of the Myers-Perry black holes is given by 
%
\begin{eqnarray}
ds^2\,&=&\,-dt^2+\frac{M}{\Sigma}(dt-j_{\phi}\sin^2\theta d\phi -j_{\psi}\cos^2\theta \psi)^2 \notag \\
&&+(r^2+j_{\phi}^{2})\sin^2\theta d\phi^2 +(r^2+j_{\psi}^{2})\cos^2\theta d\psi^2 
+\frac{\Sigma}{\Delta}dr^2 +\Sigma d\theta^2, 
\end{eqnarray} 
%
where 
%
\begin{equation}
\Sigma= r^2+j_{\phi}^{2}\cos^2\theta +j_{\psi}^{2}\sin^2\theta 
\end{equation}
and
\begin{equation}
\Delta = r^2\left(1+\frac{j_{\phi}^2}{r^2}\right)\left(1+\frac{j_{\psi}^2}{r^2}\right) -M.
\end{equation} 
%
Introducing the new coordinate defined by $\rho = 1/r$ and taking the 
conformal factor as $\Omega=\rho^{2}$, we can 
compute the multipole moments of the Myers-Perry black holes. The results are 
%
\begin{gather}
P\,=\,\frac{M}{2}\,,\,P_{a}\,=\,0\,,\, Q\,=\,-(j_{\phi}^{2}-j_{\psi}^{2})M \\
J\,=\,0\,,\,J_{\phi}\,=\,j_{\phi}M\,,\,J_{\psi}\,=\,-j_{\psi}M\, \\
J^{\phi}_{ab}\,=\,J^{\psi}_{ab}\,=\,0.
\end{gather} 
%
Contrasted with the Schwarzschild black hole case, the rotating black holes have non-zero 
mass quadrupole moments, 
which are contributions from the rotations. To measure the deviation of other  black 
object solutions from the Myers-Perry black holes, it is better to 
define the reduced mass quadrupole moments as 
%
\begin{equation}
Q^{\text{red}}\,=\,Q\,+\,\frac{J_{\phi}^{2}-J_{\psi}^{2}}{2P}. 
\end{equation} 
%
It is chosen so that the reduced mass quadrupole moments of the Myers-Perry black holes vanishes, that is, 
\begin{equation}
Q^{\text{red}}_{\text{\tiny MP}}=0.  
\end{equation}

\subsection{Black ring with single angular momentum} 

Next, we compute the multipole moments of black ring with single angular momentum. 
For the case of black ring with two angular momenta, see appendix. B.
The metric is given by \cite{Emparan:2008eg} 
%
\begin{eqnarray}
ds^2\,&=&-\frac{F(y)}{F(x)}\left( dt-CR\frac{1+y}{F(y)}d\phi \right)^2 \notag \\
&&+\frac{R^2}{(x-y)^2}F(x)\left[-\frac{G(y)}{F(y)}d\phi^2 -\frac{dy^2}{G(y)}
 +\frac{dx^2}{G(x)}+\frac{G(x)}{F(x)}d\psi^2 \right], \label{BRmetric}
\end{eqnarray} 
%
where
%
\begin{equation}
F(\xi)\,=\,1+\lambda\xi\,,\,G(\xi)\,=\,(1-\xi^2)(1+\nu\xi),
\end{equation} 
%
%
\begin{equation}
C=\sqrt{\lambda (\lambda -\nu )\frac{1+\lambda}{1-\lambda}}
\end{equation}
%
and the parameter range is $0<\nu\leq \lambda<1$. For the regular black ring solution which 
has no conical singularities, the parameters $\lambda$ and $\nu$ must satisfy the relation 
as $\lambda=2\nu/(1+\nu^2 )$. In the following, we will consider this regular black ring solution.

For the computation of the multipole moments, it is better to 
introduce the coordinate $(\rho,\theta)$ defined by 
%
\begin{equation}
x\,=\,-1+\frac{2R^2(1-\lambda)}{1-\nu}\rho^2\cos^2\theta~~{\rm and}~~
y\,=\,-1-\frac{2R^2(1-\lambda)}{1-\nu}\rho^2\sin^2\theta, 
\end{equation} 
%
and take conformal factor as $\Omega=\rho^2$. 
Then, the multipole moments of black ring with single angular momentum are evaluated as 
%
\begin{gather}
P\,=\,\frac{R^2\lambda}{1-\nu}\,,\,P_{a}\,=\,0 \\
Q\,=\,-2R^4\lambda\frac{(1+\lambda-3\nu+\lambda\nu)}{(1-\nu)^3} \\
J\,=\,0\,,\,J_{\phi}\,=\,\frac{2R^3\sqrt{\lambda(1+\lambda)(\lambda-\nu)}}{(1-\nu)^2}\,,
\,J_{ab}\,=\,0
\end{gather} 
%
and the reduced quadrupole moment becomes 
%
\begin{equation}
Q^{\text{red}}_{\text{\tiny BR}}\,=\,-\frac{2R^4\nu(1-\lambda)^2}{(1-\nu)^3} \leq 0.
\end{equation} 
%
As in static cases, the reduced quadrupole moments of black ring solutions have 
always non-positive value. In the appearance of naked singularity with $\lambda=0,1$ or 
in the Myers-Perry limit or flat metric limit, the reduced quadrupole moment becomes to be zero.  

\subsection{Classification issue}

\begin{figure}[tbp]
 \begin{minipage}[t]{0.45\hsize}
  \begin{center}
   \includegraphics[width=\hsize,clip]{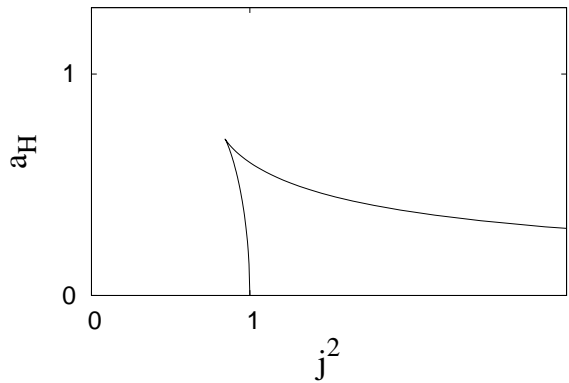}
   \caption{Normalized horizon area $a_{H}=\sqrt{2\nu(1-\nu)}$ vs. spin. Even if given the spin $j\sim\nu$, 
            fat ring ($\nu<1/2$) and thin ring ($\nu>1/2$) are not distinguished. }       
   \label{fig:five}
  \end{center}
 \end{minipage}
\end{figure}

As shown above, the Myers-Perry black hole and black ring solutions with single angular momentum are 
classified by (reduced) mass quadrupole moments completely. 
When one specifies the ADM mass and angular momentum of black ring solutions, there are two 
different solutions, thin ring and fat ring solutions in a certain of parameter region. 
To see this, it is useful to introduce the quantity $j^{2}$ as
%
\begin{equation}
j^{2}\,\equiv \,\frac{27}{32}\frac{J_{ADM}^{2}}{M_{ADM}^{3}}\,=\,\frac{(1+\nu)^3}{8\nu}.
\end{equation} 
%
Regular black ring solution has the two independent parameters $R$ and $\nu$ satisfying 
$0<R$ and $0<\nu<1$ as in Eq. (\ref{BRmetric}). If $M_{ADM}$ and $J_{ADM}$ of black ring are given, 
we can compute the value of $j^{2}$ and determine the parameter $\nu$. 
However, in the range $27/32<j^2<1$, it is known that there are 
two different solutions, that is, thin ring ($1>\nu>1/2$) and fat ring solutions ($\nu<1/2$) 
shown in Fig. \ref{fig:five}. Then, even if we assume the horizon topology of $S^{1}\times S^{2}$, we 
cannot specify the solution only by $M_{ADM}$ and $J_{ADM}$. This is well-known fact. 
%
\begin{figure}[tbp]
 \begin{minipage}[t]{0.45\hsize}
  \begin{center}
   \includegraphics[width=\hsize,clip]{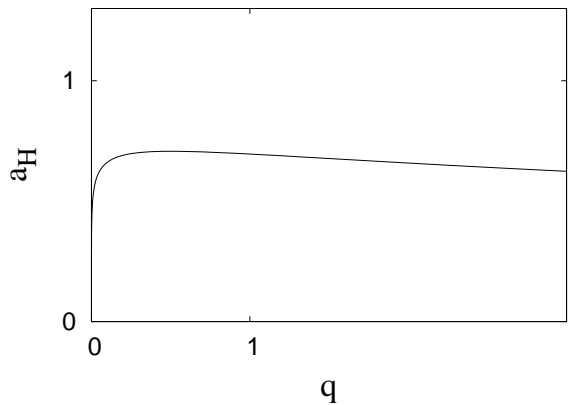}
   \caption{Normalized horizon area vs. quadrupole moment. If given the quadrupole $q$, $\nu$ is 
            completely determined, that is, thin or fat is specified. }
   \label{fig:six}
  \end{center}
  \end{minipage}
\end{figure}
%

Then, one wonders if one can distinguish these two solutions (thin or fat) by the reduced quadrupole 
moment. The answer is yes. To see this, we define $q$ as
%
\begin{equation}
q\,=\,\frac{-4Q^{\text{red}}_{\text{\tiny BR}}}{P^{2}}=\frac{2(1-\nu)^3}{\nu}.
\end{equation} 
%
The relation between the normalized area of the event horizon $a_{H}$ \cite{Emparan:2008eg} 
and $q$ is shown in Fig. \ref{fig:six}. 
From Fig. 4,  one can see that we can determine the parameter $R$ and $\nu$ if the 
reduced quadrupole moments $Q^{\text{red}}$ and mass monopole $P$ of black ring 
solutions are given. Thus, black ring 
solutions are completely specified by the mass monopole and the reduced quadrupole moments.  

Here we comment on the multipole moments for black objects with {\it multiple} horizons, e.g., 
black Saturn 
\cite{Elvang:2007rd} and orthogonal 
black di-ring solutions \cite{Izumi:2007qx,Elvang:2007hs}, although this is beyond the scope of our 
current paper. 
As shown in appendix C, these solutions all have non-vanishing reduced quadrupole moment.
Regular black Saturn solution which has no conical singularity has four independent parameters. 
Hence, by tuning these parameters, black Saturn 
has the same (reduced) quadrupole and angular dipole moments of black ring solution.  
This means that these multiple horizon solutions cannot be classified 
only by monopole and quadrupole moments. Hence, we need the information about the 
higher multipole moments for the complete classification as we pointed out in static cases. 
For the details, see appendix C.

\section{Summary and Discussion}

In this paper, we have defined mass and angular multipole moments in five dimensional stationary 
space-times. It is known that black holes and black ring with or without rotations cannot be distinguish 
by mass monopole (ADM mass) and angular dipole moment (ADM angular momentum). 
But, we could show that 
these black objects with the single horizon can be classified by introducing the (reduced) quadrupole 
moment. These moments are well defined in the center-of-mass gauge. In static cases, 
we could see that the 
mass quadrupole moments capture the existence of finite spacelike rods and 
the quadrupole 
moments detects the deviation of the topology of the event horizon from sphere. 
As seen in the previous section, 
this interpretation is valid for the reduced quadrupole moment in stationary cases.

Let us discuss the remaining works. When one wants to classify black objects with multiple 
horizons, we need the gauge independent definition of higher multipole moments. 
However, our current definition of multipole moments higher than quadrupole moments are not 
gauge invariant even in center-of-mass gauge. Therefore, we have to define the multipole moment carefully. Since the 
computation itself of the multipole moment defined here is hard task, we would guess that 
the improvement is also hard one. This might be done by additional some extra terms in our current 
definition of the multipole moments. It is also interesting to extend the definition of multipole
moments to non-vacuum cases like Einstein-Maxwell system or so. 

It is known that 
the metric is determined 
completely if we specify the all mass and angular multipole moments in four dimensional space-times \cite{Beig:1972,Beig:1981,Kundu:1980rn}. 
The method of the proof of this theorem 
does not hold in five dimensions because the fact that the Weyl tensor for $h_{ab}$ trivially vanishes 
plays a key role of the proof in four dimensions. Therefore, it is rather non-trivial if the metric 
can be determined only by mass and angular multipole moments. To investigate this, 
it may be useful to use the rod structure. 
If we can show that the parameters of the rod structure can be constructed only by mass and 
angular multipole moments,
the five dimensional black objects with single or multiple horizons in stationary 
and two rotational symmetric space-times will be classified only by mass and angular multipole moments.  
This is also our future work.

\section*{Acknowledgment}
We thank Shunichiro Kinoshita, Ryosuke Mizuno, Norihiro Tanahashi and Takahiro Tanaka for useful discussions. 
KT is supported by JSPS Grant-Aid for Scientific Research. 
TS is partially supported by Grant-Aid for Scientific Research from Ministry of Education, Science,
Sports and Culture of Japan (Nos.~21244033,~21111006,~20540258 and 19GS0219). 
This work is also supported by the Grant-in-Aid for the Global COE Program 
gThe Next Generation of Physics, Spun from Universality 
and Emergence" from the Ministry of Education, Culture, Sports, Science and Technology (MEXT) 
of Japan.

\appendix
\section{field equations}

In this appendix, we describe the key ingredient behind the definition of the mass multipole 
moments. We first write down the vacuum Einstein 
equations $R^{(5)}_{ab}=0$ for the metric of Eq. (\ref{metric}) as 
%
\begin{gather}
\hat{D}^2\lambda\,=\,\frac{1}{2\lambda}\hat{D}_{a}\lambda \hat{D}^{a}\lambda 
-\frac{1}{2\lambda}\hat{\sigma}_{ab}\hat{\sigma}^{ab} \label{lapse} \\ 
\hat{D}^{a}\hat{\sigma}_{ab}\,=\,-\frac{3}{2\lambda ^2}\hat{\sigma}_{ab}\hat{D}^{a}\lambda \label{shift} \\
\hat{R}_{ab}\,=\,\frac{1}{2\lambda}\hat{D}_{a}\hat{D}_{b}\lambda -\frac{1}{4\lambda}\hat{D}_{a}
\lambda\hat{D}_{b}\lambda- \frac{1}{4\lambda^2}(\hat{h}_{ab}\hat{\sigma}_{mn}\hat{\sigma}^{mn}-
\hat{\sigma}_{am}\hat{\sigma}_{b}^{~m}),
\end{gather}
%
where $\hat{D}$ and $\hat{R}_{ab}$ are the covariant derivative and Ricci tensor 
for $\hat{h}_{ab}$ respectively, and $\hat{\sigma}_{ab}=\hat{\epsilon}_{abcde}\xi^{c}\hat{\nabla}^{d}\xi^{e} $. 
Using the function $\hat{P}\,=\,1-\sqrt{\lambda}$, we rewrite Eq. (\ref{lapse}) as 
%
\begin{equation}
\left( \hat{D}^2 -\frac{\hat{R}}{6}\right)\hat{P} \,=\,-\frac{\hat{\sigma}_{ab}\hat{\sigma}^{ab}}{8}(2-\hat{P}), \label{lapse2}
\end{equation}
%
where $\hat{R}$ is the Ricci scalar of $\hat h_{ab}$. 
We can regard Eq. (\ref{shift}) as the Maxwell equations on the $t=\text{const.}$ hypersurfaces 
and $\hat{\sigma}_{ab}$ as the fields strength. From the conformal invariance of 
the Maxwell equation in four dimensions, 
we can suppose that the  Maxwell field $\hat{\sigma}_{ab}$ is conformal invariant 
$\hat{\sigma}_{ab}=\sigma_{ab}$ under the conformal completion of Eq. (\ref{conf-trans}). 
Then, the conformal transformation transforms Eq. (\ref{lapse2}) into 
%
\begin{equation}
\left(D^{2}-\frac{R}{6}\right) P\,=\, \Omega^{2}\frac{\sigma_{ab}\sigma^{ab}P}{8} - \frac{\Omega}{4}\sigma_{ab}\sigma^{ab}, \label{lapse3}
\end{equation}
%
where $P=\Omega^{-1}\hat{P}$. We can regard Eq. (\ref{lapse3}) as a Poisson-like equation 
with a certain of regular source. At spatial infinity $\Lambda$, it becomes 
\begin{equation}
D^{2}P\hat{=}0.
\end{equation} 
Therefore, it is natural to 
define multipole moment using $P$ and $\sigma_{ab}$.

\section{Black ring solutions with two angular momenta}

In the main text, we focused on the black objects with single angular momentum mainly. 
This is because we wanted the argument to be compact as possible as we can. 
In this appendix, we compute the multipole moments for black ring with two angular momenta. 
The metric of black ring with two angular momenta is given by \cite{Pomeransky:2006bd} 
%
\begin{eqnarray}
ds^{2}\,=\,\frac{H(y,x)}{H(x,y)}(dt\,+\,\Omega)^2 \,+\,\frac{F(x,y)}{H(y,x)}d\phi^{2} &+&2\frac{J(x,y)}{H(y,x)}d\phi d\psi 
\,-\,\frac{F(y,x)}{H(y,x)}d\psi^{2} \notag \\
&&-\frac{2k^2H(x,y)}{(x-y)^2(1-\nu)^2}\left( \frac{dx^{2}}{G(x)}-\frac{dy^{2}}{G(y)}  \right),
\end{eqnarray}
%
where
%
\begin{equation}
\Omega\,=\, -\frac{2k\lambda\sqrt{(1+\nu )^2-\lambda^{2}}}{H(y,x)}
\left\{ (1-x^2)y\sqrt{\nu}d\psi +\frac{(1+y)}{1-\lambda +\nu}(1+\lambda-\nu +x^{2}y
\nu (1-\lambda -\nu ) +2\nu x(1-y))d\phi \right\},
\end{equation}
%
and
%
\begin{eqnarray}
G(x)&=&(1-x^2)(1+\lambda x +\nu x^2) \\
H(x,y)&=&1+\lambda^2 -\nu^2 +2\lambda\nu (1-x^2)y+2\lambda x(1-\nu^2 y^2)+\nu x^2 y^2(1-\lambda^2 -\nu^2) \\
J(x,y)&=&\frac{2k^2(1-x^2)(1-y^2)\lambda\sqrt{\nu}}{(x-y)(1-\nu)^2}\{ 1+\lambda^2 -\nu^2 +2(x+y)\lambda\nu -xy\nu(1-\lambda^2 -\nu^2) \} \\
F(x,y)&=&\frac{2k^2}{(x-y)^2(1-\nu)^2}\left[ G(x)(1-y^2)\{ ((1-\nu)^2-\lambda^2 )(1+\nu )+y\lambda ( 1-\lambda^2 +2\nu -3\nu^2 )  \}  \right. \notag \\
&&~~~~~~~~~~~~~~~~~~~~~~~~~+G(y)\left( 2\lambda^2 +x\lambda \{ (1-\nu)^2+\lambda^2 \} + x^2\{ (1-\nu)^2 -\lambda^2 \}(1+\nu) \right.  \\
&&~~~~~~~~~~~~~~~~~~~~~~~~~~~+x^3\lambda (1-\lambda^2 -3\nu^2 +2\nu^3) -x^4 \nu(1-\nu) (-1+\lambda^2 +\nu^2) \left.\right)\left.\right] \notag .
\end{eqnarray}
%
The parameter ranges are $0<\nu<1$, $2\sqrt{\nu}<\lambda<1+\nu$. Regular black ring solution with two 
angular momenta have three independent parameters $\nu$, $\lambda$ and $k$.  

For computing the multipole moments, we introduce the new coordinate $(\rho,\theta)$ defined by 
%
\begin{equation}
x\,=\,-1+\frac{4k^2(1-\lambda +\nu)}{1-\nu}  \rho^2\cos^2\theta \,,\,
y\,=\,-1-\frac{4k^2(1-\lambda +\nu)}{1-\nu}  \rho^2\sin^2\theta, 
\end{equation}
%
and use the conformal factor of $\Omega=\rho^2$. Then, the multipole moments are computed as 
%
\begin{gather}
P\,=\,\frac{4k^2\lambda}{(1-\lambda +\nu)}\,,\,P_{a}\,=\,0, \\
Q\,=\,-\frac{16\lambda k^4(1-5\nu -5\nu^2 +\nu^3 -8\lambda \nu +3\lambda^2( 1+\nu ) )}{(1-\nu)^2(1-\lambda +\nu)^2} \\
J_{\psi}\,=\,\frac{16k^3\lambda\sqrt{\nu}\sqrt{(1+\nu)^2-\lambda^2}}{(1-\nu)^2(1-\lambda +\nu)}\,,\,
J_{\phi}\,=\,\frac{8k^3\lambda(1+\lambda -6\nu +\lambda\nu +\nu^2)\sqrt{(1+\nu)^2-\lambda^2}}{(1-\nu)^2(1-\lambda +\nu)^2}.
\end{gather}
%
Then the reduced quadrupole moment becomes 
%
\begin{equation}
Q^{\text{red}}\,=\,-\frac{8\lambda k^4(1-\lambda +\nu )}{(1-\nu)^2}.
\end{equation}
%
As in one rotational case, black ring has a negative value for the mass quadrupole moment. Thus, in two rotational case,
Myers-Perry black hole and black ring with two angular momenta can be classified by the mass quadrupole moment.

\section{Cases with multiple horizons}

In this appendix, we consider the cases with multiple horizons. 
In the main text, we focused on single horizon cases and we could show that 
space-times are uniquely specified by the multipole moments up to the quadrupole components. 
We can show that it is not true for the cases with multiple horizons. This section will 
be useful for future study or comparison with single horizon cases. 

\subsection{Static cases}

%
\begin{figure}[tbp]
 \begin{tabular}{cc}
 \begin{minipage}[t]{0.45\hsize}
  \begin{center}
   \includegraphics[width=\hsize,clip]{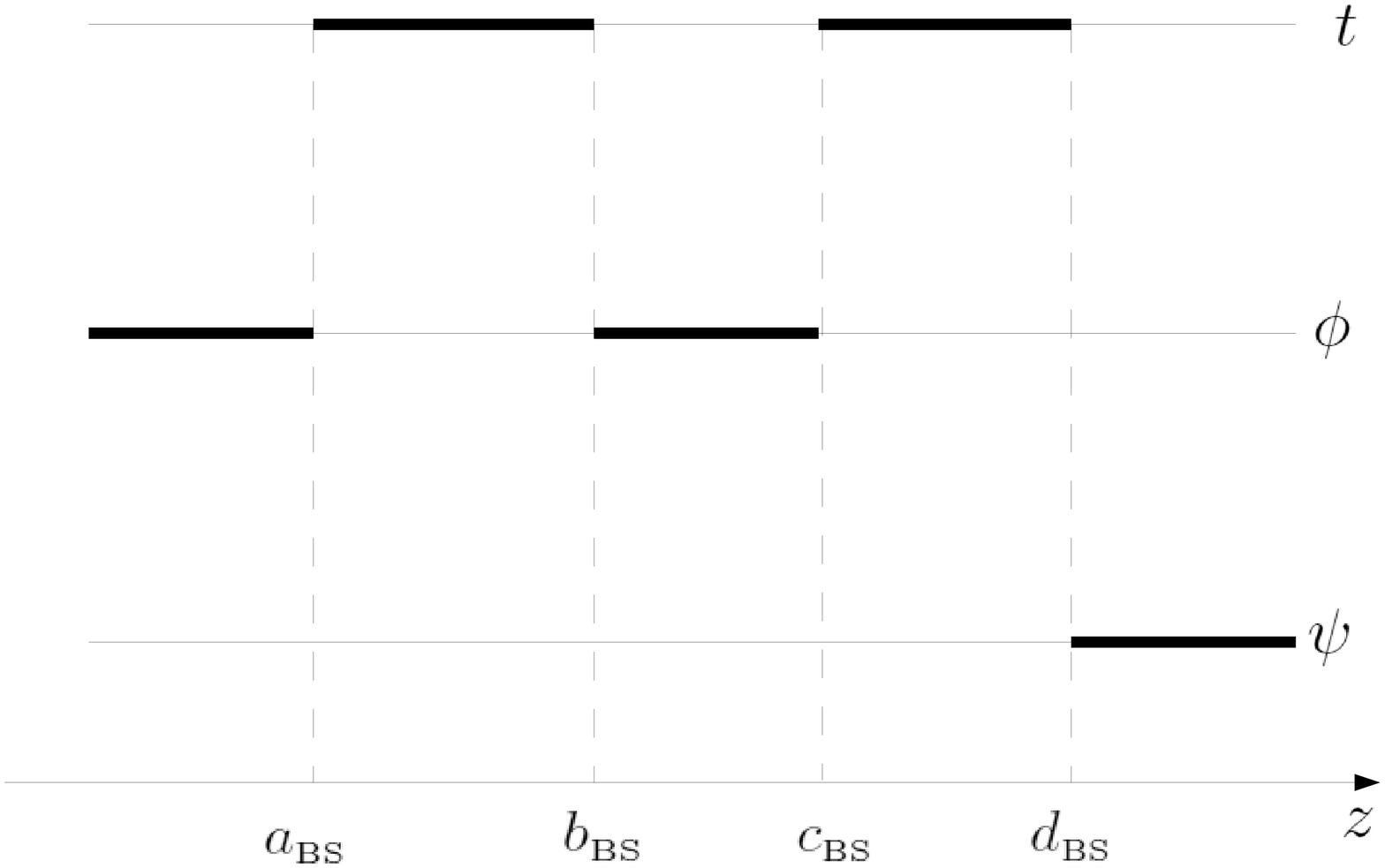}
   \caption{Rod structure of static black Saturn} 
   \label{fig:three}
  \end{center}
  \end{minipage}
 \begin{minipage}[t]{0.45\hsize}
  \begin{center}
   \includegraphics[width=\hsize,clip]{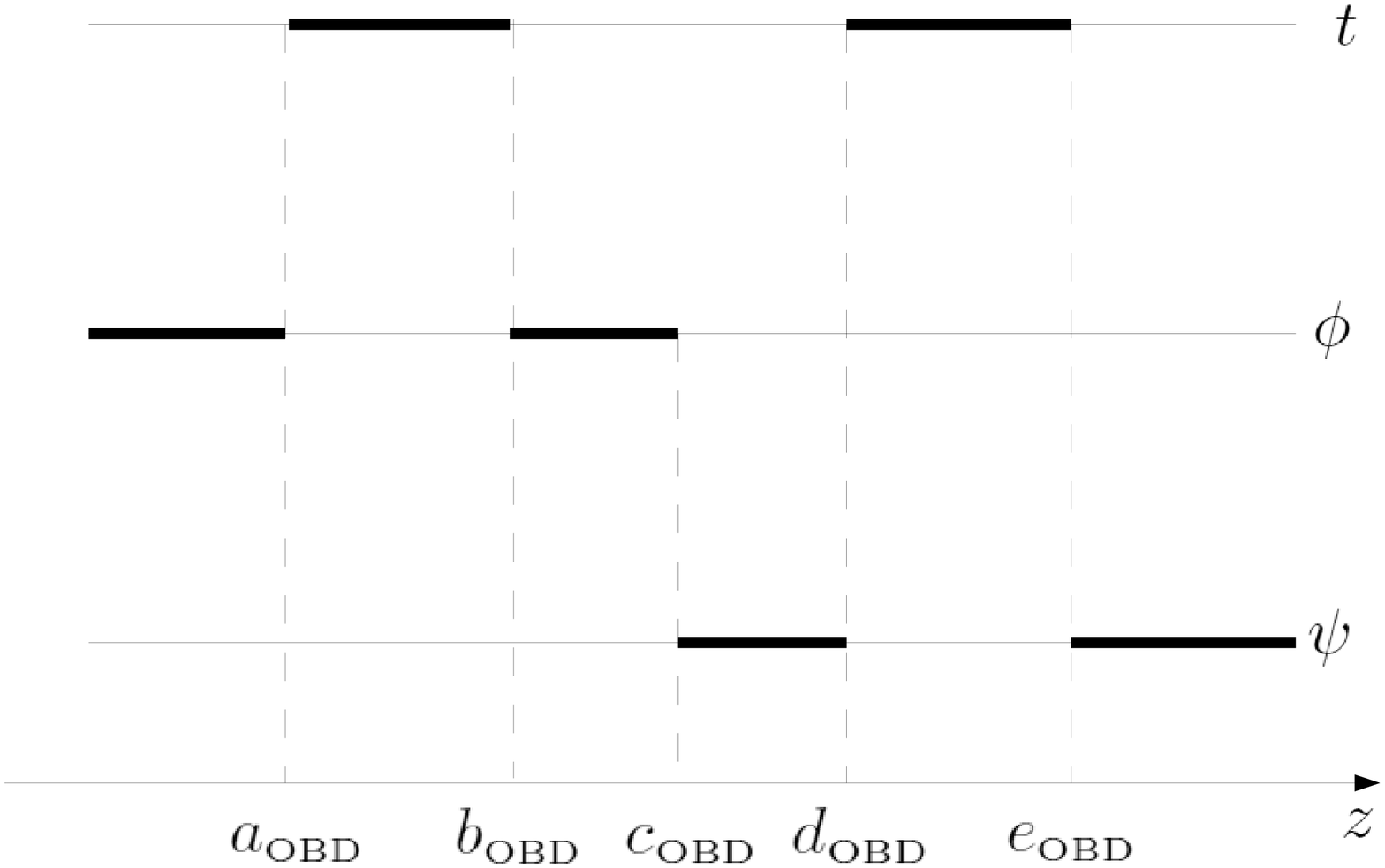}
   \caption{Rod structure of static orthogonal black di-ring}
   \label{fig:four}
  \end{center}
  \end{minipage}
 \end{tabular}
\end{figure}
%

Here we 
compute multipole moments for static black Saturn solution 
and static orthogonal black di-ring solution (these rod structures shown in Figs. \ref{fig:three} 
and \ref{fig:four}). The multipole moments are 
%
\begin{gather}
P\,=\,(b_{\text{\tiny BS}}-a_{\text{\tiny BS}})+(d_{\text{\tiny BS}}-c_{\text{\tiny BS}})\,,\,P_{a}\,=\,0 \\
Q=\,-4(b_{\text{\tiny BS}}-a_{\text{\tiny BS}})(c_{\text{\tiny BS}}-b_{\text{\tiny BS}})  
\end{gather} 
%
for static black Saturn, and
%
\begin{gather}
P\,=\,(b_{\text{\tiny OBD}}-a_{\text{\tiny OBD}})+(d_{\text{\tiny OBD}}
-c_{\text{\tiny OBD}})\,,\,P_{a}\,=\,0 \\
Q\,=\,Q_{1}+Q_{2} \\
Q_{1}\,=\,-4(b_{\text{\tiny OBD}}-a_{\text{\tiny OBD}})
(c_{\text{\tiny OBD}}-b_{\text{\tiny OBD}})\,
,\,Q_{2}\,=\,4(e_{\text{\tiny OBD}}-d_{\text{\tiny OBD}})
(d_{\text{\tiny OBD}}-c_{\text{\tiny OBD}})  
\end{gather} 
%
for static orthogonal black di-ring. 

We can see that in the quadrupole moments of the static orthogonal black di-ring, $Q_{1}$ is 
quadrupole moments of black ring in the $\phi$ rotational plane, and $Q_{2}$ is the quadrupole 
moment of black ring 
in the $\psi$ rotational plane. Hence, 
monopole and quadrupole moments are ``linear" moments. 
Only by monopole and quadrupole 
moments, we cannot distinguish static black ring, black Saturn and black di-ring. 
Static black Saturn solutions have three independent parameters, for example, 
${b_{\text{\tiny BS}}-a_{\text{\tiny BS}}}$, $c_{\text{\tiny BS}}-b_{\text{\tiny BS}}$ and
$d_{\text{\tiny BS}}-c_{\text{\tiny BS}}$. By tuning these parameters, static black 
Saturn can have the same mass monopole and quadrupole moments as static black ring's.
As in static orthogonal black di-ring solutions, we can do same thing because 
its solution has four independent parameters. 
That is, there are several different solutions with same $P$ and $Q$. 
This result suggests that higher multipole moments are needed to 
classify all these solutions.

\subsection{Stationary cases}

\begin{figure}[tbp]
 \begin{tabular}{cc}
 \begin{minipage}[t]{0.45\hsize}
  \begin{center}
   \includegraphics[width=\hsize,clip]{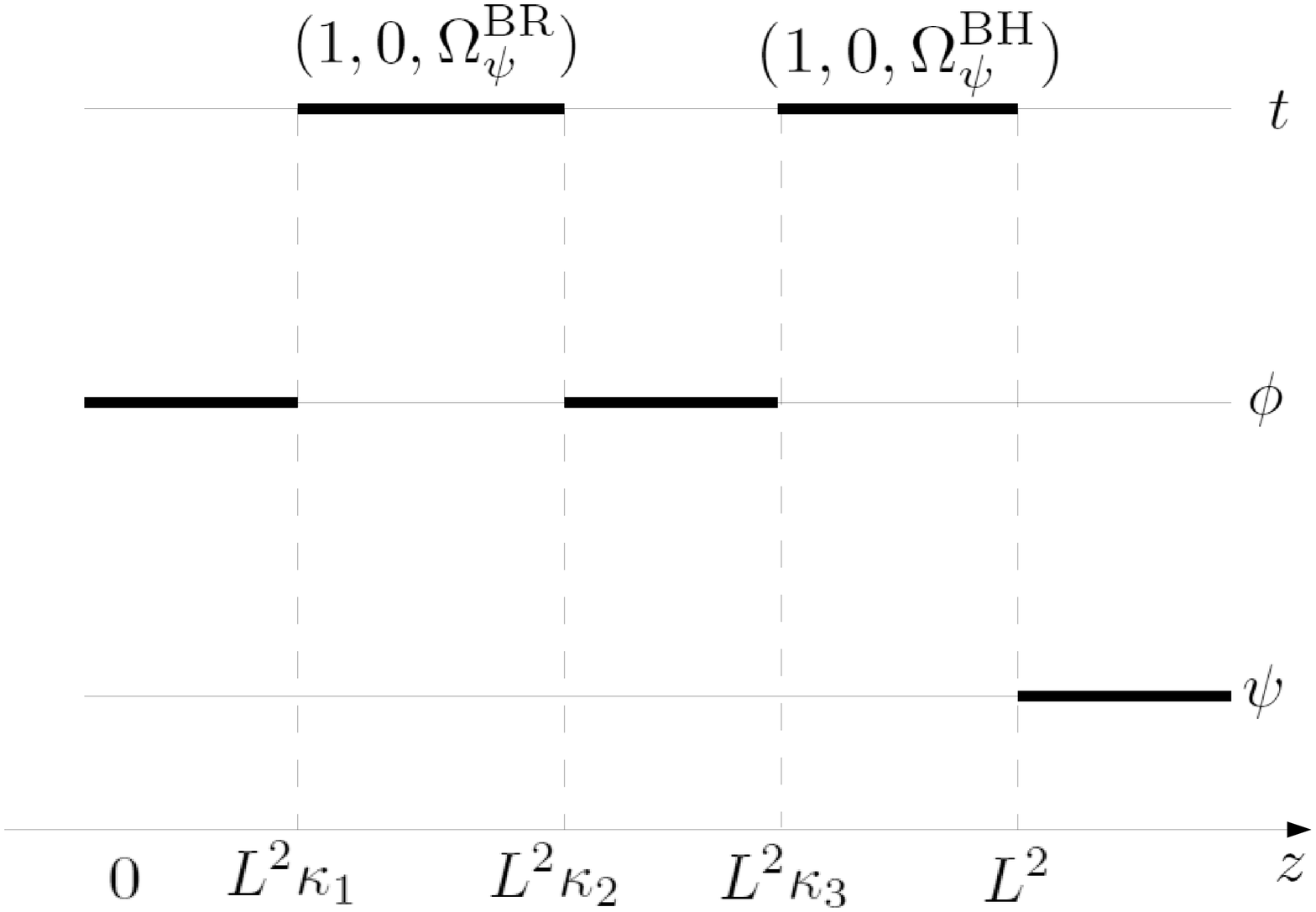}
   \caption{Rod structure of black Saturn} 
   \label{fig:seven}
  \end{center}
  \end{minipage}
 \begin{minipage}[t]{0.45\hsize}
  \begin{center}
   \includegraphics[width=\hsize,clip]{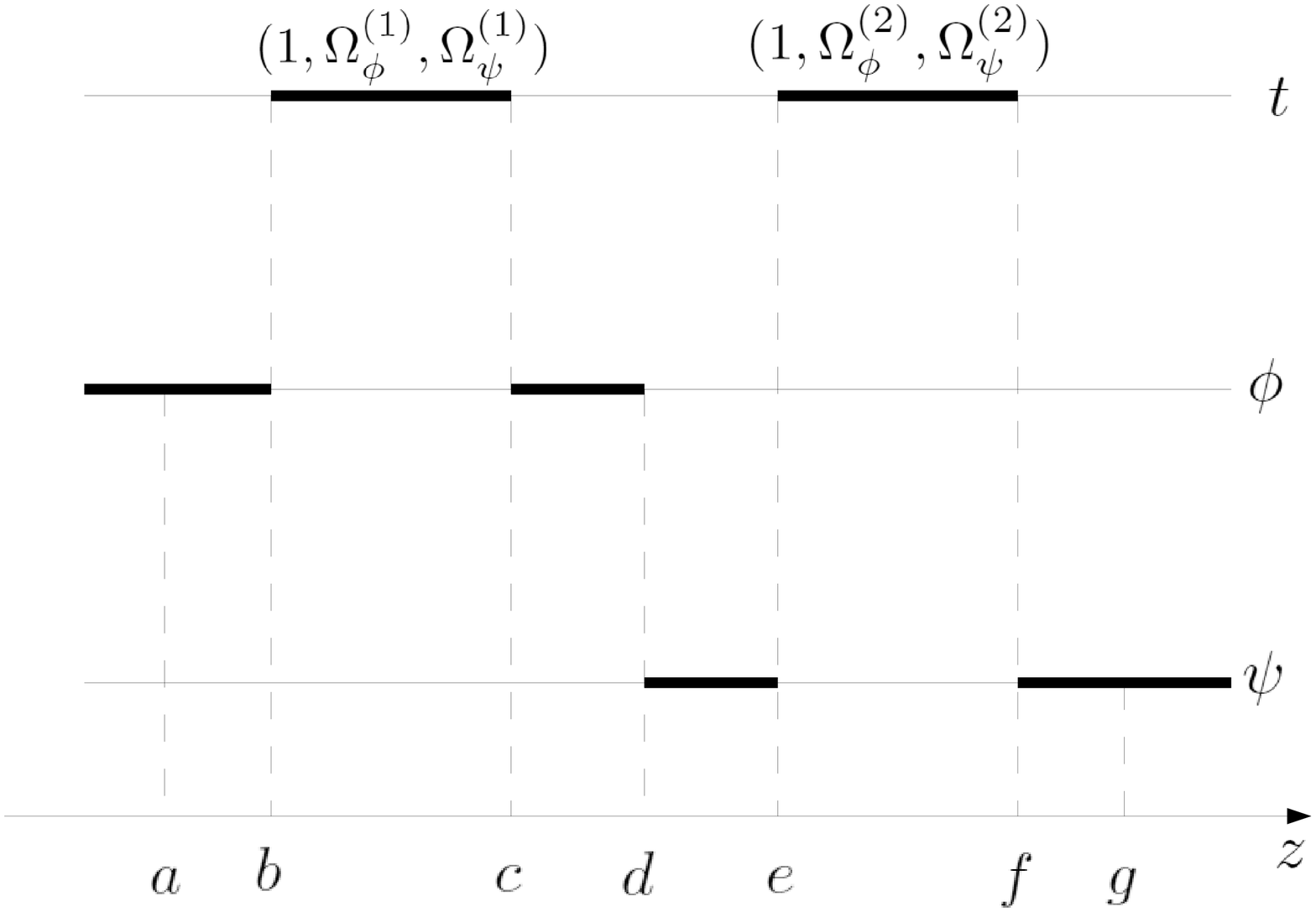}
   \caption{Rod structure of orthogonal black di-ring}
   \label{fig:eight}
  \end{center}
  \end{minipage}
 \end{tabular}
\end{figure}

Here, we compute the multipole moments for the stationary solutions with multiple horizons, 
black Saturn \cite{Elvang:2007rd} and orthogonal black di-ring \cite{Izumi:2007qx}. 
Since the explicit form of the metric is complicated, we show only 
the rod structure of the solutions (See Figs. \ref{fig:seven} and \ref{fig:eight}).    
%
%
The metric of stationary and two rotational symmetric solutions in 
five dimensions can be written by Weyl form as
%
\begin{eqnarray}
ds^{2}\,=\,G_{AB}dx^{A}dx^{B} +e^{2\nu}(dR^{2}+dz^{2}),
\end{eqnarray} 
%
where $x^{A}=(t,\phi,\psi)$. The rod structure is described by the zero point of $\det G_{AB}$ and their 
direction $\xi^{A}$ determined from $G_{AB}\xi^{A}=0$ at $R=0$. For computing the multipole moments, 
it is better to introduce the new coordinate $(\rho,\theta)$ defined through the relation 
%
\begin{eqnarray}
R\,=\,\frac{1}{2\rho^{2}}\sin 2\theta\,,\,z\,=\,\frac{1}{2\rho^{2}}\cos 2\theta,
\end{eqnarray} 
%
and we choose the conformal factor as $\Omega=\rho^{2}$. 

The rod structure of black Saturn solution is shown in Fig. \ref{fig:seven}. The angular velocities are 
given by  
%
\begin{eqnarray}
\Omega^{\text{BH}}_{\psi}&=& \frac{1}{L}(1+\kappa_{2}\bar{c}_{2})\sqrt{\frac{\kappa_{2}\kappa_{3}}{2\kappa_{1}}}
\frac{\kappa_{3}(1-\kappa_{1})-\kappa_{1}(1-\kappa_{2})(1-\kappa_{3})\bar{c}_{2}}
{\kappa_{3}(1-\kappa_{1})+\kappa_{1}\kappa_{2}(1-\kappa_{2})(1-\kappa_{3})\bar{c}_{2}^{2}} \\
\Omega^{\text{BR}}_{\psi}&=& \frac{1}{L}(1+\kappa_{2}\bar{c}_{2})\sqrt{\frac{\kappa_{1}\kappa_{3}}{2\kappa_{2}}}
\frac{\kappa_{3}-\kappa_{2}(1-\kappa_{3})\bar{c}_{2}}
{\kappa_{3}-\kappa_{3}(\kappa_{1}-\kappa_{2})\bar{c}_{2}+\kappa_{1}\kappa_{2}(1-\kappa_{3})\bar{c}_{2}^{2}},
\end{eqnarray} 
%
for each event horizons, that is, for the central black hole and outer black ring. 
Note that the following regularity condition is imposed
%
\begin{eqnarray}
\bar{c}_{2}\,=\,\frac{1}{\kappa_{2}}\left[\epsilon\frac{\kappa_{1}-\kappa_{2}}{\sqrt{\kappa_{1}(1-\kappa_{2})
(1-\kappa_{3})(\kappa_{1}-\kappa_{3})}
}\right],
\end{eqnarray} 
%
where $\epsilon=1 (-1)$ for $\bar{c}_{2}>-\kappa_{2}^{-1} (\bar{c}_{2}<-\kappa_{2}^{-1})$. 
Thus, black Saturn 
solutions has four independent parameters $L$, $\kappa_{1}$, $\kappa_{2}$ and $\kappa_{3}$. 

After some length calculations, the mass multipole moments for black Saturn are computed as 
%
\begin{eqnarray}
P &=& L^{2}\frac{\kappa_{3}(1-\kappa_{1}+\kappa_{2})-2\kappa_{2}\kappa_{3}(\kappa_{1}-\kappa_{2})\bar{c}_{2}+\kappa_{2}
[\kappa_{1}-\kappa_{2}\kappa_{3}(1+\kappa_{1}-\kappa_{2})]\bar{c}_{2}^{2}}{\kappa_{3}(1+\kappa_{2}\bar{c}_{2})^2} \\
P_{a}&=&0 \\
Q &=& 
\frac{4L^{4}}{(\kappa^{2}_{3}(1+\kappa_{2}\bar{c}_{2})^{4})}
(\kappa^{2}_{3}[(\kappa_{2}-\kappa_{1})(\kappa_{2}-\kappa_{3})-\kappa_{3}]
+2\kappa_{2}\kappa^{2}_{3}[1+2(\kappa_{2}-\kappa_{1})(\kappa_{2}-\kappa_{3})-\kappa_{3}]\bar{c}_{2}
\notag \\
&&~~~
+\kappa_{2}\kappa_{3}[3\kappa_{2}\kappa_{3}\{ 1+2(\kappa_{2}-\kappa_{1})(\kappa_{2}-\kappa_{3})\}-\kappa_{1}(1-\kappa_{1}+\kappa_{2}+2\kappa_{3})]\bar{c}_{2}^{2}
\notag \\
&&~~~
+2\kappa_{2}^{2}\kappa_{3}[\kappa_{2}\kappa_{3}\{ 2(\kappa_{2}-\kappa_{3})(\kappa_{2}-\kappa_{1})+\kappa_{3}\} +\kappa_{1}(1+\kappa_{1}-\kappa_{2}-2\kappa_{3})]\bar{c}_{2}^{3}
\notag \\
&&~~~~
+\kappa_{2}^{2}\left[\kappa_{2}^{2}\kappa_{3}^{2}\{ (\kappa_{2}-\kappa_{1})(\kappa_{2}-\kappa_{3})-(1-\kappa_{3})\} -\kappa_{1}\{ \kappa_{1}-\kappa_{2}\kappa_{3}(3+\kappa_{1}-\kappa_{2}-2\kappa_{3})\} \right] \bar{c}_{2}^{4}).
\end{eqnarray} 
%
And the angular multipole moments are also computed as 
%
\begin{eqnarray}
J^{\phi} &=& 0 \\
J_{\phi} &=& 
\frac{4L^{3}}{\kappa_{3}(1+\kappa_{2}\bar{c}_{2})}\sqrt{\frac{\kappa_{2}}{\kappa_{1}\kappa_{3}}}
\left[\kappa^{2}_{3}-\kappa_{3}\bar{c}_{2}[(\kappa_{1}-\kappa_{2})(1-\kappa_{1}+\kappa_{3})+\kappa_{2}(1-\kappa_{3})]
\right.\notag\\
&&~~~~~~~~~~~~~~~~~~~~~~~~~~~~~
+\kappa_{2}\kappa_{3}\bar{c}_{2}^{2}[(\kappa_{1}-\kappa_{2})(\kappa_{1}-\kappa_{3})+\kappa_{1}(1+\kappa_{1}-\kappa_{2}
-\kappa_{3})]\notag\\
&&~~~~~~~~~~~~~~~~~~~~~~~~~~~~~\left.
-\kappa_{1}\kappa_{2}\bar{c}^{3}_{2}[\kappa_{1}-\kappa_{2}\kappa_{3}(2+\kappa_{1}-\kappa_{2}-\kappa_{3})]
\right] \\
J^{\phi}_{ab} &=& 0.
\end{eqnarray} 
%

Next we consider the orthogonal black di-ring solution. 
The rod structure is shown in Fig. \ref{fig:eight}. The angular velocities are given by 
%
\begin{gather}
\Omega^{(1)}_{\phi}\,=\,-j_{1}\frac{(g-e)(g-f)}{2(g-c)^{2}(g-a)}\,,\,
\Omega^{(1)}_{\psi}\,=\,-j_{2}\frac{(e-a)^{2}}{2(c-a)(f-a)(g-a)} \\
\Omega^{(2)}_{\phi}\,=\,-j_{1}\frac{(g-f)}{2(g-a)(g-d)}\,,\,
\Omega^{(2)}_{\psi}\,=\,-j_{2}\frac{(d-a)^{2}}{2(f-a)(g-a)}.
\end{gather} 
%
The labeling of $(1), (2)$ specifies which event horizons we consider. 
The regularity conditions is also imposed as 
%
\begin{gather}
j_{1}^{2}\,=\,2\frac{(g-a)(g-c)^{2}(g-d)}{(g-b)(g-e)(g-f)}\,,\,j_{2}^{2}\,=\,2\frac{(b-a)(c-a)(f-a)(g-a)}{(d-a)(e-a)^{2}} \\
(a-d)(a-f)(a-g)(b-e)(b-g)(c-d)(c-e)(c-f)\,=\,(a-e)^{2}(b-d)^{2}(b-f)^{2}(c-g)^{2} \\
(g-d)(g-b)(g-a)(f-c)(f-a)(e-d)(e-c)(e-b)\,=\,(g-c)^{2}(f-d)^{2}(f-b)^{2}(e-a)^{2}.
\end{gather} 
%
Note that there is the gauge freedom of $z\rightarrow z+\text{const.}$. 
Thus, the orthogonal black di-ring has the four free parameters. 

The mass multipole moments for orthogonal black di-ring are computed as
%
\begin{gather}
P\,=\,(c-a)+(g-e) \\
P_{a}\,=\,0 \\
Q\,=\,Q_{\phi}-Q_{\psi}
\end{gather} 
%
where
%
\begin{gather}
Q_{\phi}\,=\,-4(c-a)[(b-a)+(d-c)+(f-e)]\,,\,Q_{\psi}\,=\,-4(g-e)[(e-d)+(c-b)+(g-f)].
\end{gather} 
%
We can interpret the $Q_{\phi}$ and $Q_{\psi}$ as the quadrupole moment of the black ring 
in the $\phi$- and $\psi$-rotational planes, respectively. 

The angular momentum multipole moment are
%
\begin{gather}
J^{\phi,\psi}\,=\,0 \\
J_{\phi}\,=\,2j_{1}(d-a)\,,\,J_{\psi}\,=\,-2j_{2}(g-b)(g-e)(g-f) \\
J^{\phi,\psi}_{ab}\,=\,0.
\end{gather} 
%

As in static cases, regular black objects with multiple horizons have non-trivial 
quadrupole moments.
These solutions with multiple horizons have independent parameters more than three. 
Hence, to classify these solutions, it is necessary to evaluate 
higher multipole moments such as mass $2^{4}$-pole moments.



\end{document}